\documentclass[usegraphicx]{mn2e}
\def\H0{{\it H}$_0$}
\def\Ms{{\it M}$_\odot$}
\def\Ls{{\it L}$_\odot$}
\def\q0{{\it q}$_0$}

\def\ergps{erg~s$^{-1}$}
\def\kmpspMpc{km~s$^{-1}$~Mpc$^{-1}$}
\def\Ms{{\it M}$_\odot$}

\def\nH{$N_{\rm H}$\thinspace} 

\def\psqcm{cm$^{-2}$}
\def\ergpspsqcm{erg~cm$^{-2}$~s$^{-1}$}

\def\phpspsqcm{ph\thinspace s$^{-1}$\thinspace cm$^{-2}$}

\def\pcubcm{cm$^{-3}$}
\def\ergcmps{erg\thinspace cm\thinspace s$^{-1}$}

\def\ts{\thinspace}

\title[Fe K line in Arp220] 
{Fe K emission in the ultraluminous infrared galaxy Arp220}
\author[K. Iwasawa et al]
{\parbox[]{6.5in} {K. Iwasawa$^1$, D. B. Sanders$^2$, A. S. Evans$^3$, N. Trentham$^1$, G. Miniutti$^1$ and H.W.W. Spoon$^4$}\\
  \\
  $^1$Institute of Astronomy, Madingley Road, Cambridge CB3 0HA\\
  $^2$Institute for Astronomy, University of Hawaii, 2680 Woodlawn Drive, Honolulu, HI 96822, USA\\
 $^3$Department of Physics and Astronomy, Stony Brook University, Stony Brook, NY 11794-3800, USA\\
$^4$Astronomy Department, Cornel University, Ithaca, NY 14853, USA\\
} \date{}

\begin{document}

\maketitle
\begin{abstract}
  Prominent Fe K$\alpha$ line emission is detected in the XMM-Newton
  spectrum of the ultraluminous infrared galaxy Arp220. The centroid
  of the line is found at an energy of 6.7 keV and the equivalent
  width of the line is $EW\sim 1.9$ keV (at $3.5 \sigma$
  significance). A few other spectral features are found at various
  degrees of significance in the lower energy range on a hard 2.5--10
  keV continuum ($\Gamma\sim 1$). The large EW of the Fe K line poses
  a problem with interpreting the hard X-ray emission as integrated
  X-ray binary emission. A thermal emission spectrum with a
  temperature of $kT\sim 7$ keV modified by absorption of \nH $\simeq
  3\times 10^{22}$\psqcm, can describe the 2.5--10 keV continuum shape
  and the Fe K emission.  A hot bubble that is shocked internally in a
  starburst region would have a similar temperature and gives a good
  explanation for the observed X-ray properties with a high star
  formation rate. An ensemble of radio supernovae in a dense
  environment, as suggested from VLBI imaging, could be another
  possibility, if such powerful supernovae are produced continuously
  at a high rate. However, the apparent lack of emission from X-ray
  binaries is incompatible with the high supernova rate ($\sim 2$ SNe
  yr$^{-1}$) required by both interpretations.  Highly photoionized,
  low-density gas illuminated by a hidden Compton-thick AGN is a
  possible alternative for the hard X-ray emission, which can be
  tested by examining whether radiative recombination continua from
  highly ionized Ca and Fe are present in better quality data from a
  forthcoming observation.
\end{abstract}

\begin{keywords}
  Galaxies: individual: Arp220 --- X-rays: galaxies
\end{keywords}

\section{Introduction}

Arp220 is the nearest ultraluminous infrared galaxy (ULIG), at $z =
0.018$ (corresponding to $d = 74$ Mpc for $H_{0} = 75$\kmpspMpc), with
$> 95$ per cent of its total bolometric luminosity emitted at
infrared/submillimetre wavelengths [$L_{\rm ir}(8-1000\mu m) \simeq
1.3\times 10^{12}$\Ls] (Soifer et al. 1984; Sanders et al 1988, 2003).
It is an advanced merger system involving two relatively large spiral
disks, as evidenced by the detection in the optical of two large,
faint, crossed tidal tails (Joseph \& Wright 1985; Sanders et al
1988), with two nuclei separated by $\approx 1$ arcsec ($\approx 300$
pc), as determined from images in the radio (e.g. Becklin \&
Wynn-Williams 1987; Norris 1988; Baan \& Haschick 1995) and
near-infrared (e.g., Graham et al 1990; Scoville et al 2000) bands.
High resolution CO observations imply an extreme nuclear molecular gas
concentration ($\sim 3 \times 10^9$\Ms\ at radii $< 300$pc
corresponding to a mean H$_2$ density of $10^4$-$10^5$ \pcubcm) (e.g.,
Scoville et al 1986; Solomon, Radford, Downes 1990; Scoville et al
1998; Sakamoto et al 1999).  This is probably a consequence of disk
gas being funnelled into the central regions of both galaxies during
the merger (e.g., Barnes \& Hernquist 1992). There is evidence for a
galactic-scale outflow, as H$\alpha $ and soft X-ray nebulae with a
size of $\sim 30$ kpc (Armus, Heckman \& Miley 1990; Heckman et al
1996) suggest, which is thought to be driven by a super-starburst (but
see Colina, Arribas \& Clements 2004 for a merger shock
interpretaion). The optical spectrum is LINER-like (Veilleux, Kim \&
Sanders 1999), most likely due to shock-heated interstellar gas by the
galactic wind (e.g., Taniguchi et al 1999), while the double nucleus
region is heavily obscured by dust.

Arp220 is often used as a nearby template of luminous star-forming
galaxies at high redshift, thus verifying the true dominant power
source in Arp220 continues to be of great importance.  There is still
no convincing direct evidence, from radio to hard X-ray wavelengths,
for an active nucleus in Arp220.  The general assumption has been that
the bolometric luminosity of Arp220 is mostly powered by a starburst,
as perhaps best represented by a view based on ISO mid-infrared
spectral characteristics (e.g. the PAH feature at 7.7 $\mu $m --
Genzel et al 1998; Sturm et al 1996).  However, the interpretation of
the ISO data for ULIGs, and Arp220 in particular, has recently been
revised by Spoon et al (2004), who took ice absorption into account
and concluded that the PAH component is rather weak with only moderate
obscuration. They inferred that a major power source, whether a super
star cluster or an AGN, must be deeply enshrouded in dust, similar to
an interpretation previously proposed by Dudley \& Wynn-Williams
(1997).  A few authors have suggested that a hidden QSO could be a
major power source for Arp220 (e.g., Sanders et al 1988; Haas et al
2001). If an energetically significant AGN is present in Arp220, it
must be Compton-thick (e.g., Rieke 1988). Hard X-rays constrain the
lower bound of the absorbing column density to be \nH $\sim 10^{25}$
\psqcm\ (Iwasawa et al 2001).  The lack of cold reflection
characterised by a 6.4 keV Fe K line means that the covering factor of
the obscuring matter has to be close to unity.

Recent studies of nearby elliptical galaxies suggests that all large
spheroids contain a supermassive black hole (SMBH: e.g., Richstone et
al 1998).  Given that the host galaxy of Arp220 already appears to
have relaxed into an elliptical-like $r^{1/4}$-law profile (e.g.
Wright et al 1990), it seems reasonable to speculate that Arp220 may
also contain a SMBH, perhaps even two, since the two observed nuclei
have yet to merge.  Two SMBH have recently been discovered in the
luminous infrared galaxy NGC6240 (Komossa et al 2003), which is also
an advanced merger system similar to Arp220. Given the detection of an
enormous nuclear concentration of molecular gas in Arp220 there is
plenty of material in the nuclear region to feed, and therefore build
a SMBH, while also simultaneously suppressing the amount of direct
emitted radiation from the nuclear source. This is not an implausible
scenario, and a sensitive hard X-ray observation would have a better
chance to catch the faintest sign of such a hidden AGN than
observations at longer wavelengths, as sensitive hard X-ray
observations of NGC4945 (Iwasawa et al 1993; Done et al 1996;
Guainazzi et al 2001) and NGC6240 (Iwasawa \& Comastri 1998; Vignati
et al 1999; Ikebe et al 2000) have shown.

We present here our analysis and interpretations of the public archive
XMM-Newton data on Arp220, with which a strong Fe K line is detected
for the first time. This will hopefully be followed by a longer observation
that should be scheduled sometime during AO-3.

\section{Observations and data reduction}

Arp220 was observed with XMM-Newton on 2002, August 11, and 2003,
January 15. The two XMM-Newton observations were carried out in the
Full Window mode and, when combined together, provide a useful
exposure time of 19{\ts}ks. We only use the EPIC pn data because of
its high sensitivity in the Fe K band. The data analysis presented
here was performed using the latest version of the standard analysis
package, SAS 6.0. Single and double events from the detector were
selected and the data reduction was carried out following the standard
procedure.

\section{Results}

\subsection{The XMM-Newton EPIC spectrum}

% Fig. 1 --- XMM spectrum

\begin{figure}
\centerline{\includegraphics[width=0.37\textwidth,angle=270,
    keepaspectratio='true']{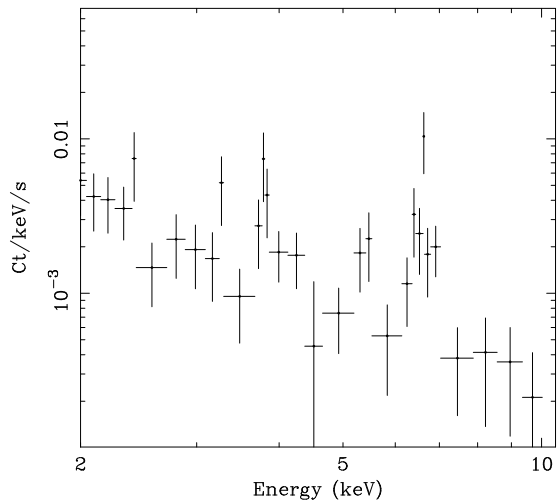}} 
\caption{
The 2--10 keV band XMM-Newton EPIC pn spectrum of Arp220. }
\end{figure}

The hard X-ray ($\geq 3$ keV) emission in Arp220 is only resolved at
the resolution of the Chandra X-ray Observatory (Clements et al 2002;
and see Section 3.2), and is point-like when viewed with the
XMM-Newton telescope, although a much larger extension of soft X-ray
emission is clearly resolved. As we focus on the hard X-ray emission,
the spectral data are taken from a circular region of a
$25$ arcsec radius, enough to collect most of the hard X-ray
photons. The detailed location of the hard X-ray source is shown in
Section 3.2.

The EPIC pn spectrum is shown in Fig. 1. The soft X-ray emission below
$\sim 2 $keV is mostly due to the extended nebula, within which a wide
range of spectral variations are present between regions, as revealed
by the Chandra data (McDowell et al 2003). 
The data analysis presented here is restricted to
the energy range 2.5--10 keV to focus on the hard X-ray emission around
the double nucleus.

A prominent Fe K$\alpha $ line is detected at $3.5\sigma $
significance (Fig. 1). The centroid of the feature is found at
$6.72^{+0.05}_{-0.05}$ keV (in the rest frame; the errors represent
the 90 per cent confidence region for one parameter of interest),
indicating that FeXXV is the major line component.  Fitting a single
gaussian gives a line flux of $1.7^{+0.8}_{-0.8}\times 10^{-6}$
\phpspsqcm. The corresponding equivalent width (EW) against the
underlying continuum is $1.9\pm 0.9$ keV.  The inclusion of the
gaussian line in the fitted model reduces $\chi^2$ by $\simeq 10$ (23
degrees of freedom). The single gaussian fit does not require
statistically significant broadening, but a multiple line complex in
the 6.5--7 keV range would be more likely.

There are suggestions of other spectral features at 3.3, 3.9, 4.2 and
5.5 keV with their detection being at $2.6\sigma $ or lower. They all
have possible identifications with a highly ionized plasma. Because of
their low significance, we only give results of spectral fitting with
narrow gaussians for those emission features in Table 1, and will not
use them as critical materials for further discussion. However, we
point out that they would be an important key to constrain the origin
of the hard X-ray emission, once their detections are confirmed by
better quality data. Especially, radiative recombination continua are
spectral features unique to photoionized gas (Section 4.5).

% Table 1.  Emission lines
\begin{table}
\begin{center}
  \caption{ Emission line features in the 3--10 keV band. Results are
    obtained from fitting each spectral feature with a narrow
    gaussian, and the centroid energy is corrected for the galaxy
    redshift ($z=0.018$).  The fit with these gaussian lines gives
    $\chi^2_{\nu}=1.0$ for 17 degrees of freedom. The errors quoted
    are the 90 per cent confidence region for one parameter of
    interest. RRC stands for radiative recombination continuum, which
    is only relevant for photoionized gas. $\dagger $The line flux
    limits are obtained by fitting with the energies
    fixed for ArXVIII and CaXX, respectively.}
\begin{tabular}{cccc}
$E$ & $I$ & $EW$ & ID \\
keV & $10^{-7}$\phpspsqcm & keV & \\[5pt]
$3.32\dagger $ & $3.7^{+5.7}_{-3.7}$ & 0.20 & ArXVIII, SXIV RRC \\
$3.87^{+0.06}_{-0.05}$ & $8.9^{+6.7}_{-5.6}$ & 0.55 & CaXIX \\
$4.11\dagger $ & $2.1^{+9.5}_{-2.1}$ & 0.17 & CaXX \\
$5.50^{+0.10}_{-0.26}$ & $4.7^{+5.5}_{-4.1}$ & 0.45 & CaXX RRC \\
$6.72^{+0.10}_{-0.12}$ & $17.4^{+8.1}_{-7.8}$ & 1.85 & FeXX-FeXXVI \\
%$6.57^{+0.10}_{-0.10}$ & $10.6^{+6.2}_{-6.5}$ & 1.28 & FeXX-FeXXII??? \\
%$7.02^{+0.09}_{-0.32}$ & $7.7^{+6.0}_{-5.5}$ & 1.00 & FeXXVI \\
\end{tabular}
\end{center}
\end{table}

The 2.5--10 keV continuum is very hard: fitting a power-law modified
only by Galactic absorption (\nH $= 4\times 10^{20}$\psqcm) gives
$\Gamma = 1.2^{+0.4}_{-0.7}$.  The total 2--10 keV flux is $1.1\times
10^{-13}$\ergpspsqcm. This value is in agreement with the value
obtained from the Chandra observation (Clements et al 2002), but is
smaller than the BeppoSAX MECS value $1.8\times 10^{-11}$\ergpspsqcm.
As pointed out by Clements et al (2002), the MECS aperture contains
two hard X-ray sources to the south, and the discrepancy with BeppoSAX
is probably due to contamination from these sources. The
corresponding 2--10 keV luminosity is $7\times 10^{40}$\ergps, and
$L_{\rm 2-10keV}/L_{\rm ir}\simeq 1.5\times 10^{-5}$.

The possibility that these contaminating sources in the XMM-Newton
beam emit the iron line emission can be ruled out. The nucleus of
Arp220 is the only significant source within the beam in the narrow
band (6--7 keV) Chandra image (see Section 3.2). Given the large EW of
the line (which means the iron line emission to dominates the 6--7 keV
band), the Arp220 nucleus is certainly the iron line source.

\subsection{Chandra imaging}

%Fig .4 --- Chandra-HST overlay

\begin{figure}
\centerline{\includegraphics[width=0.37\textwidth,angle=270,
%    keepaspectratio='true']{nicmos-chandra.ps}} 
    keepaspectratio='true']{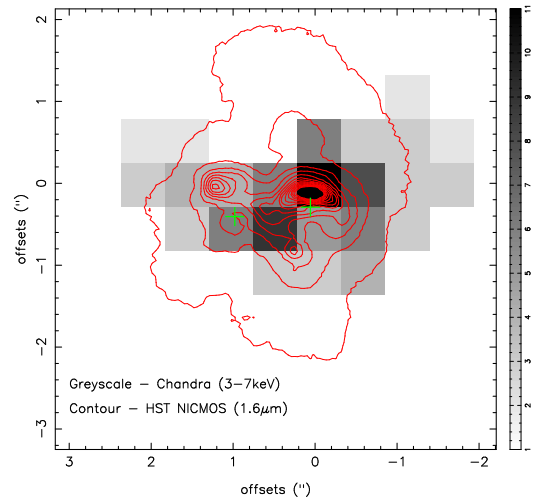}} 
\caption{
  The Chandra 3--7 keV image of the double nucleus region of Arp220
  overlaid by the HST NICMOS 1.6 $\mu $m contour map (Scoville et al
  2000). The positions of the double radio nuclei are indicated with
  plus symbols in green.}
\end{figure}

As reported previously, the hard X-ray emission is slightly extended
but concentrated around the double nucleus (Clements et al 2002). Most
of the 3--7 keV emission comes from radii within 0.7 kpc and very
little originates beyond 1 kpc. Fig. 2 shows the Chandra 3--7 keV band
image with the HST NICMOS 1.6$\mu $m image superposed. The astrometry
of the Chandra image has been corrected using the latest attitude
correction file, and the HST image has been registered in the manner
discussed in Scoville et al (1998). In this updated registration, the
western nucleus (radio and near-infrared) now coincides with the 3--7
keV peak. Note the positions of the Eastern (radio) nucleus and the
1.6 $\mu$m peak, the displacement of which is presumably due to strong
obscuration (see Scoville et al 2000 for details on the near-infrared
extinction in this region). The centroid of the 6--7 keV band
emission, for which only 12 counts are detected and which is
presumably mostly due to Fe K line emission (see also Clements et al
2002), may be slightly displaced from the Western nucleus to the East,
but this is not conclusive. The Chandra spectrum of the hard X-ray
source is consistent with the spectral model with gaussian lines
fitted to the XMM-Newton spectrum in the 2--5 keV range. At higher
energies, a comparison becomes difficult as the efficiency of
Chandra declines sharply.

%\section{Interpretation}

\section{The origin of the hard X-ray emission}

\subsection{X-ray binary emission}

It is generally assumed that the 2--10 keV emission in starburst
galaxies is dominated by integrated emission from X-ray binaries, and
its luminosity has been found to be correlated with star formation
rate indicators, e.g., infrared luminosity (Nandra et al 2002;
Ranalli, Comastri \& Setti 2003; Grimm et al 2003). There may be
non-thermal inverse Compton scattered emission of infrared photons by
relativistic electrons (e.g., Moran \& Lehnert 1997), but its
contribution is probably minor.  In Arp 220, the hard X-ray emission
is resolved with Chandra and its extension coincides roughly with the
dense molecular disk (e.g., Sakamoto et al 1999) in shape. However,
the detection of the strong Fe K line readily rules out X-ray binaries
as a major source of the 2--10 keV emission because of the spectral
incompatibility.

If the infrared luminosity from Arp220 is entirely due to a starburst,
the implied star formation rate is $\approx 200$\Ms\ yr$^{-1}$
(Kennicutt 1998). With this high star formation rate, emission from
X-ray binaries -- which is estimated by assuming only high-mass X-ray
binaries\footnote{In a young starburst system, as low mass X-ray
  binaries are not yet formed, it is appropriate to consider only
  high-mass X-ray binaries (Persic \& Rephaeli 2002).}, and by
following Franceschini et al (2003; see also Persic \& Rephaeli 2002;
Persic et al 2004) -- should dominate, or in fact, exceed the observed
luminosity by more than one order of magnitude, even if absorption of
the order of $10^{22}$-$10^{23}$\psqcm\ in \nH\ is taken into account.
The lack of X-ray binary emission appears to be a general problem with
a starburst interpretation for Arp220, given the good correlation
between the 2--10 keV X-ray binary emission and the infrared
luminosity claimed for star forming galaxies (e.g., Ranalli, Comastri
\& Setti 2003; Grimm et al 2003; Persic et al 2004).

In the following subsections, we discuss two possible origins of the Fe K
line in the context of a starburst, based on the thermal emission
model for the observed X-ray spectrum. Both explanations require as
high as a star formation rate mentioned above, therefore the lack of
the X-ray binary emission remains to be a problem.

\subsection{Thermal emission model}

With the Fe K line centred at 6.7 keV, it seems tempting to interpret the hard
X-ray emission as thermal emission from hot gas associated with a
starburst. Fitting the 2.5--10 keV data with the collisionally ionized
plasma spectra computed by the MEKAL code (e.g., Kaastra 1992) gives
the following results: The temperature of the gas implied from the fit is
$7.4^{+5.4}_{-3.1}$ keV. The absorption column density is not well
constrained, but the likely value for the best-fit temperature is
$3\times 10^{22}$\psqcm. The metallicity, which is primarily
determined by the Fe K line strength, is found to be
$2.2^{+3.2}_{-1.4}$ solar (the solar abundance table by Anders \&
Grevesse 1989 is used here).

The thermal emission model accounts for the continuum shape and the Fe
K feature (with $\chi^2=21.5$ for 17 degrees of freedom), but would
leave the possible lower energy features unexplained.  Since the
interstellar gas in a starburst region is expected to be enriched by
core-collapse supernovae (e.g., Type II SNe), which produce
substantial $\alpha$-elements but a relatively small amount of iron.
The twice-solar Fe metallicity required to explain the Fe K feature is
already large, although it may not be surprising for a region with
intense star formation (e.g., Fabbiano et al 2004). A non-solar
abundance ratio between $\alpha $-elements and Fe, as expected for
chemical enrichment by Type II SNe, might be required, if the
detection of the high ionization Ar and Ca features are confirmed by
higher quality data. If confirmed, the CaXIX He-$\alpha$ at 3.9 keV
(Table 1) would not be compatible with the single temperature thermal model,
because with a temperature of 7 keV, calcium ions are mostly CaXX.

\subsection{A hot bubble in a starburst region}

According to the starburst-driven superwind model, a hot bubble of
internally shocked wind material with a temperature of several keV
will form in the starburst region (Chevalier \& Clegg 1985), which
eventually breaks away to drive a galactic-scale outflow (e.g.,
Tomisaka \& Ikeuchi 1988). Although such a hot bubble is expected not
to be radiative because of its rarefied interior, and is indeed rarely
observed directly in starburst galaxies (e.g., Suchkov et al 1994;
Strickland \& Stevens 2000; Hoopes et al 2003)\footnote{The presence
  of such hot gas has been reported in M82 by Griffiths et al (2000)
  and in NGC253 by Pietsch et al (2001), but for NGC253,
  photoionization by AGN is proposed by Weaver et al (2002).}, the
temperature of $kT\approx 7$ keV implied from the above thermal
emission model matches the prediction. In fact, the observed
temperature and luminosity agree roughly with the predicted values
from the wind solution by Chevalier \& Clegg (1985) with the mass and
energy input rates estimated by Starburst99 (Leitherer et al 1999) for
the star formation rate of 200 \Ms\ yr$^{-1}$ (see below) and a high
thermalization efficiency ($\simeq 1$). 

As the Chandra image shows, most of the hard X-ray emission in Arp220
comes from within $\approx 0.7$ kpc of the double nucleus. Assuming the thermal
plasma is uniformly distributed within a sphere with a radius of 0.7
kpc, the mean gas density and the total gas mass are derived to be
$n_{\rm gas}\simeq 0.6$ \pcubcm\ and $M_{\rm gas}\simeq 2\times
10^7$\Ms, respectively. The thermal energy contained in the volume is
estimated to be $\approx 3.5\times 10^{56}$ erg. The bolometric
luminosity of the gas of $\approx 1.6\times 10^{41}$\ergps\ means that
the cooling time of the gas is $\sim 7\times 10^7$ yr. This value is
insensitive to the non-solar abundance ratio, because, at a
temperature of $kT\sim 7$ keV, the cooling is dominated by
bremsstrahlung continuum rather than line emission.  

With the star formation rate of 200 \Ms\ yr$^{-1}$, the mass injection
rate from OB stars is $\sim 30$\Ms\ yr$^{-1}$, assuming a Salpeter IMF
and a mass loss rate of $10^{-5}$ \Ms, typical for each massive star
(see also Elson, Fall \& Freeman 1989; Heckman et al 1990).
Starburst99 also gives a mass injection rate of $\sim 50$\Ms\
yr$^{-1}$ both from stellar winds and supernovae, assuming continuous
star formation and a starburst age $>10^7$ yr. The above IMF is
assumed to have an upper mass cut-off at 100 \Ms. If the IMF is
truncated at 30 \Ms, the estimate is reduced by a factor of 2 or less.
However, given the cooling time, the expected mass input is sufficient
to supply and maintain the hot gas.

At a temperature of 7 keV, the sound speed is $c_{\rm s}\approx 800$
km s$^{-1}$. So, the sound-crossing time over the radius of the hard
X-ray emitting region (0.7 kpc), is $8\times 10^5$ yr. Since the sound
crossing time is much shorter than the cooling time, the hot gas,
which will not be static because of its high pressure ($\sim 10^{-8}$
dyne cm$^{-2}$), will drive a superwind.

In terms of the energetics of the gas, heating by SNe is sufficient to
counterbalance the radiative loss. As discussed in the following
section, the expected supernova rate is $\sim 2$ SNe yr$^{-1}$,
implying an energy injection rate of $\sim 6\times 10^{43}$ \ergps (or
Starburst99 gives the mechanical luminosity of $\sim 1\times 10^{44}$
\ergps). The required fraction of the SN energy that goes to heating
is very small $\eta_{\rm h}\sim 3\times 10^{-3}$. Therefore, the bulk
of the energy has to escape in the form of a wind -- given the
moderate soft X-ray nebula luminosity -- without depositing the energy
onto the galactic medium to radiate. This hot bubble interpretation
appears entirely plausible, apart from the X-ray binary problem
mentioned above.

\subsection{Luminous radio supernovae}

An alternative origin for the thermal emission is an ensemble of
supernovae. Several radio knots distributed over the nuclear region
have been imaged with VLBI, and are interpreted as very luminous
compact radio supernovae taking place in a dense environment (Smith et
al 1998).  They can be luminous X-ray sources, as Type IIn SNe
(Benetti et al 1995) like SN1986J (Houck et al 1998), SN1988Z (Fabian
\& Terlevich 1996) and SN1995J (Fox et al 2000) have X-ray
luminosities of $10^{40}$--$10^{41}$\ergps. Their X-ray spectra seem
hard enough to match the hard-band spectrum of Arp220, i.e.,
temperatures of 3--10 keV in $kT$ when fitted with a thermal emission
model (as summarised in Fox et al 2000), and at least SN1986J shows
evidence for a strong Fe K line at 6.7 keV (Houck et al 1998). With a
star formation rate of 200 \Ms\ yr$^{-1}$, the expected supernova rate
is $\sim 2$ SNe yr$^{-1}$, as also estimated by Smith et al (1998). If
1 per cent of the total energy of each SN goes into radiation, and is
emitted at $10^{41}$\ergps, then the cooling time is $\sim 3$ yr,
which is roughly comparable to the estimated average time between SNe
(however, as the radiative efficiency might be much higher in the
dense environment, these SNe could overproduce the integrated X-ray
luminosity).  Therefore it is plausible to have multiple X-ray SNe at
the same time, and to maintain a stable hard X-ray luminosity until
most of the massive stars die out.

Whether powerful SNe like SN1986J are constantly produced in the
Arp220 nuclear region is questioned by the second VLBI observation
following the results presented in Smith et al (1998) three years
later.  Most of the radio knots have faded by 8--50 per cent (some
remain at the same brightness), but no new compact sources have
appeared (Smith et al 1999). The slow decline is not consistent with a
luminous SN model. The SN rate also has to be lowered ($\leq 0.3$ SNe
yr$^{-1}$, Smith et al 1999). Although this estimate is only for the
powerful radio SNe, X-ray luminous SNe are probably closely related.
With these facts, it is unclear whether the above explanation of the
hard X-ray emisson with powerful SNe is sustainable.

\subsection{Photoionized gas}

Photoionized gas illuminated by a hidden AGN is still a viable
alternative for Arp220.  In fact, if the possible radiative
recombination continuum (RRC) feature is real, it may be the most
likely interpretation.  A high ionization parameter $\xi =
L_{ion}/(nR^2)\sim 10^3$ \ergcmps, where $L_{ion}$ \ergps, $n$
\pcubcm, $R$ cm, are respectively the ionizing luminosity, density and
distance from the source, is implied by the line emission centred at
6.7 keV. The extension of the hard X-ray emission points to the
presence of a larger region of low density interstellar gas with
density of the order of 1-10 \pcubcm. Extended photoionized nebulae
have been found in a number of nearby Seyfert 2 galaxies, some of which
show extended Fe K emission (e.g., NGC4945, Done et al 2003; NGC4388,
Iwasawa et al 2003).

The photoionized gas in AGN appears to have ionization parameters
distributed over a wide range (e.g., Krolik \& Kriss 2001; Kinkhabwala
et al 2002). Although detailed modelling of the Arp220 hard X-ray
spectrum is beyond the scope of this paper, given the quality (and
also the spectral resolution) of the present data, our preliminary
inspection using the photoionization code, XSTAR (Kallman \& Bautista
2001) indicates log $\xi$ in the range of 2.8--3.5 is relevant to the
observed spectral features. 
%The CaXIX emission at 4 keV would be
%present at the lower end of this ionization range. 
Absorption with a
column density of the order of $10^{23}$\psqcm\ would be required for
this photoionized spectrum not to dominate the soft X-ray band and to
make the spectrum as hard as is observed.

This photoionization model predicts RRC features at 5-5.5 keV from
CaXIX,XX and at $\sim 9$ keV from FeXXV,XXVI, which should be
relatively isolated so that they could be resolved at the CCD
resolution if the data quality is improved. This could provide a
crucial test for the photoionization model when longer exposure hard
X-ray data are obtained.

Strong Fe K line complex emission, consisting of a 6.4 keV line, which
is usually the strongest as expected from FeI-XVII, and higher energy
lines of FeXXV (6.7 keV) and FeXXVI (6.97 keV), has been observed in a
number of Compton-thick Seyfert 2 nuclei (e.g., NGC1068, NGC6240,
NGC4945). It is clear that the 6.4 keV line and the higher energy
lines originate from different matter because of their large difference
in ionization stages. The narrow line width of the latter naturally
suggests that  the line emitting gas is optically thin.

The strong 6.4 keV lines seen in good quality spectra of Seyfert 2
nuclei often show a Compton shoulder, i.e. a weak redward extension to
the line core (e.g., Iwasawa et al 1997; Bianchi et al 2002; Kaspi et
al 2002; see also George \& Fabian 1991; Matt 2002 for theory). This
means that the 6.4 keV line results from reflection from optically
thick matter, which is, in the torus model for the unification scheme,
identified as the visible surface of the inner wall of the obscuring
torus (e.g., Awaki et al 1991; Ghisellini, Haardt \& Matt 1994;
Krolik, Madau \& \.Zycki 1994).

No detection of a 6.4 keV line in the Arp220 spectrum implies the lack
of reflection from optically-thick cold matter, and perhaps suggests
that the visibility of the torus inner wall is very limited.  The
conditions for an energetically significant AGN to exist in
Arp220, imposed by the hard X-ray observation, are that the obscuring
matter must have a column density larger than \nH $= 10^{25}$\psqcm\ 
with a covering factor close to unity (Iwasawa et al 2001). With the
very high covering factor, the inner wall of the obscuring torus is
capped and never visible, which could explain the lack of a 6.4 keV
line. If a tiny fraction of light from a hidden nucleus leaks through 
the heavy obscuration, the surrounding low density medium could be highly ionized to
give rise to ``hot'' Fe K line emission. 

The covering fraction, $(1-f)$, can be estimated based on the Fe K
line flux, by comparing the well-studied Compton-thick AGN, NGC1068.
NGC1068 has a large infrared excess with $L_{\rm ir}\simeq 7\times
10^{44}$ \ergps\ for an assumed source distance of 14.4 Mpc. The 6.8 keV line
flux measured in the Chandra data is $5.5\times 10^{-5}$\phpspsqcm\ 
(Young, Wilson \& Shopbell 2001; a similar flux was measured by
Iwasawa et al 1997 for the sum of FeXXV and FeXXVI in the ASCA data).
Kinkhabwala et al (2002) estimated $fL_{ion}\approx 10^{43}$\ergps, by
fitting the detailed photoionization model to the RGS data on NGC1068.
If both NGC1068 and Arp220 are powered by hidden AGN, and the hot Fe K
line emitting region in NGC1068 sees the same ionizing source, then
$fL_{ion}$ for Arp220 is $\sim 0.8\times 10^{43}$\ergps, and with
$L_{ion}\sim (1/2)L_{\rm ir}$, $f$ is estimated to be $\sim 0.4$ per
cent, which can be compared with $f\sim 3$ per cent for NGC1068
($f<10$ per cent from the Chandra image, Kinkhabwala et al 2002) under
the same assumption. Thus, reducing the opening fraction of the
obscuration further would make the photoionization model consistent
with that for NGC1068.

\section{The power source of the far-infrared emission}

While there is no doubt that a starburst is taking place in Arp220,
various mid-infrared characteristics, summarised by Spoon et al (2004,
and reference therein), the superwind luminosity inferred from the
H$\alpha $ and soft X-ray nebulae (Suchkov et al 1996; Iwasawa 1999;
also McDowell et al 2003 for an alternative interpretation of merger
shock), and the X-ray binary emission in the hard X-ray band, as
discussed above, are all unusually low relative to the far-infrared
luminosity for starburst galaxies.

The peculiarity of Arp220 is well illustrated by Spoon et al (2004),
who re-examined the ISO mid-infrared spectrum with the knowledge of a
number of ice absorption features across the band. Their spectral
analysis and the higher resolution study by Soifer et al (2002)
demonstrate that the mid-infrared emission in Arp220 consists of
diffuse PAH emission with a moderate amount of absorption and a
NGC4418-like, heavily absorbed continuum. The latter is more likely to
contribute to the far-infrared luminosity of Arp220, although the
above authors note that, if all the far-infrared luminosity is powered
by a hidden AGN, it would not even be visible at mid-infrared
wavelengths.

If, say, 10 per cent, of the bolometric luminosity ($\sim 10^{11}$\Ls
) was due to a moderately obscured starburst, the outward
characteristics of such a starburst as described above would then be
in agreement with most other starburst galaxies. For example, the soft
X-ray nebula has the luminosity of $\sim 1\times 10^{41}$\ergps,
consistent with the predicted luminosity ratio of X-ray and
starburst-powered infrared emission for a superwind nebula (log
(SX/IR) $\approx -3.5$, Leitherer \& Heckman 1999; Strickland \&
Stevens 2000). Also, the X-ray binary emission in the hard X-ray band
approaches the correlation line with the star formation rate for local
starburst galaxies (Ranalli et al 2002; Grimm et al 2002). 

If such a lower luminosity starburst was indeed the case, the rest
of Arp220's luminosity would have to be explained by something else.
Spoon et al (2004) suggested a deeply shrouded ultra dense starburst,
in addition to the less obscured starburst, which is responsible for
the diffuse PAH emission. The estimated optical depth based on their
modelling of the ISO spectrum implies that the column density to such
a star cluster would be \nH $\sim 10^{23}$\psqcm. Since X-rays at
energies higher than a few keV are transparent to this level of
obscuration, luminous hot gas, for instance, associated with the star
cluster would be observable as a hard X-ray excess, which is not
apparent in the data.

An alternative option is an even more deeply buried AGN. If the
photoionized gas interpretation was correct for the detected Fe K
line, then the presence of an AGN would be proved. It is not
straightforward to estimate the intrinsic luminosity of such an
obscured AGN without seeing its transmitted radiation: once the line
of sight Thomson depth exceeds unity (Compton thick), the visibility
of nuclear X-ray emission is largely determined by geometry, e.g., the
covering factor, distribution of the gas density etc., which are not
known. However, the comparison with NGC1068 (Section 4.4) demonstrates
that with $f\sim 0.4$ per cent, the detected Fe K line luminosity is
consistent with the assumption that all the luminosity is powered by
an AGN.  Therefore once the photoionization origin for the Fe K line is
confirmed, it would be plausible that an AGN dominates the energetics.
The radio spectrum of Arp220 must then be explained by free-free
absorption, the opacity of which should be larger than those in the
Compton-thick AGN in NGC6240 and NGC4945 (see Fig. 5 in Iwasawa et al
2001).

There are a few interesting infrared luminous objects to compare with
Arp220. NGC4418 is a luminous infrared galaxy with $L_{\rm ir}\simeq
1\times 10^{11}$\Ls, and its mid-infrared spectrum shows no PAH
features, but is shaped by silicate and various ice absorption
features (Spoon et al 2001). Very faint X-ray emission with $L_{\rm
  X}\sim 10^{39}$ \ergps\ has been detected from the centre of NGC4418
with Chandra (Maiolino et al 2003). The luminosity ratio of X-ray and
infrared emission for NGC4418 {\it is even smaller} than that of
Arp220. The hyper-luminous infrared galaxy, IRAS 00182--7112
($z=0.327$), has kinematical evidence for a superwind (Heckman et al
1990), as in Arp220, and its mid-infrared spectrum is classified as
Class 1 by Spoon et al (2002), i.e., it is NGC4418-like. A recent
XMM-Newton observation of IRAS 00182--7112 detected a Fe K line at 6.7
keV with $EW\sim 1.5$ keV on a flat continuum (K. Nandra, priv comm),
reminiscent of Arp220.

These three objects share similar properties, and if they are powered
by stars, they are starbursts for which the widely used relation
between far-infrared (or the translated star formation rate) and hard
X-ray (X-ray binary emission) luminosities does not apply. Their 2--10
keV X-ray luminosities are well below the correlation line of Ranalli
et al (2003) and Grimm et al (2003). The detection of the Fe K line,
in Arp220 in particular, means that their X-ray binary contribution to
the 2--10 keV band must be minor, making them deviate from the
correlation further. Hiding most of the X-ray binaries behind thick
obscuration required a substantial mass of gas and dust, given the
observed spatial extension (e.g., Arp220). If the formation of X-ray
binaries were suppressed, an unusually large suppressing factor ($\sim
20$ for Arp220 when compared with Grimm et al 2003) would be required.
An heavily obscured AGN may offer a more relaxed solution for the
embedded power sources for all three of these objects. For Arp220, of
course, a substantial starburst ($\sim 10^{11}$\Ls) would still be
required to explain the near-mid infrared light, superwind signatures
and radio supernovae etc.

\section*{Acknowledgements}

The XMM-Newton data presented here were obtained from the XMM-Newton
Science Archive maintained by the Science Operations Centre and the
observations were carried out in the Guaranteed-Time program (PI, B.
Aschenbach).  We thank Steve Allen, Steve Smartt, Massimo Ricotti,
Andy Fabian, and Dave Strickland for useful discussion and Paul Nandra
for information on his unpublished result. ASE was supported by NSF
grant AST 00-80881.  GM, KI and NT thank PPARC for support.


\begin{thebibliography}{}


\bibitem{} Anders E., Grevesse N., 1989, Geochim. Cosmochim. Acta, 53, 197

\bibitem{} Armus L., Heckman T.M., Miley G.K., 1990, ApJ, 364, 471

\bibitem{} Awaki H., Koyama K., Inoue H., Halpern J.P., 1991, PASJ, 43, 195

\bibitem{} Baan W.A., Haschick A.D., 1995, ApJ, 454, 745

\bibitem{} Barnes, J.E., Hernquist, L., 1992, ARAA, 30, 705

\bibitem{} Becklin, E.E., Wynn-Williams, C.G., 1987, in Star Formation in Galaxies, (Washington, D.C.: U.S. Government Printing Office), ed. C.J. Lonsdale, p643

\bibitem{} Benetti S., Bouchet P., Schwarz H., 1995, IAU Circ., 6170

\bibitem{} Bianchi S., Matt G., Fiore F., Fabian A.C., Iwasawa K., Nicastro F., 2002, A\&A, 396, 793

\bibitem{} Brinkman A.C, Kaastra J.S., van der Meer R.L.J., Kinkhabwala A., Behar E., Khan S.M., Paerels F.B.S., Sako M., 2002, A\&A, 396, 761

\bibitem{} Chevalier R.A., Clegg A.W., 1985, Nat, 317, 44

\bibitem{} Clements D.~L., McDowell J.~C., Shaked S., Baker A.~C., Borne K., Colina L., Lamb S.~A., Mundell C., 2002, ApJ, 581, 974 

\bibitem{} Colina L., Arribas S., Clements D., 2004, ApJ, 602, 181

\bibitem{} Done C., Madejski G.M., \.Zycki P.T., Greenhill L.J., 2003, ApJ, 588, 763

\bibitem{} Dudley C.C., Wynn-Williams C.G., 1997, ApJ, 488, 720

\bibitem{} Elson R.A.W., Fall S.M., Freeman K.C., 1989, ApJ, 336, 734

\bibitem{} Evans A.S., et al, 2003, AJ, 125, 2341
 
\bibitem{} Fabbiano G. et al., 2004, ApJ, 605, L21

\bibitem{} Fabian A.C., Terlevich R., 1996, MNRAS, 280, L5

\bibitem{} Fox D.W., et al., 2000, MNRAS, 319, 1154

\bibitem{} Franceschini A., et al., 2003, MNRAS, 343, 1181

\bibitem{} Ghisellini G., Haardt F., Matt G., 1994, MNRAS, 267, 743 

\bibitem{} Genzel R., et al, 1998, ApJ, 498, 579

\bibitem{} George I.M., Fabian A.C., 1991, MNRAS, 249, 352

\bibitem{} Graham, J.R., Carico, D.P., Matthews, K., Neugebauer, G., Soifer, B.T., Wilson, T.D., 1990, ApJ, 354, L5

\bibitem{} Griffiths R.E., Ptak A., Feigelson E.D., Garmire G., Townsley L., Brandt W.N., Sambruna R., Bregman J.N., 2000, Science, 290, 1325

\bibitem{} Grimm H.-J., Gilfanov M., Sunyaev R., 2003, MNRAS, 339, 793

\bibitem{} Haas M., Klaas U., M\"uller S.A.H., Chini R., Coulson I., 2001, A\& A, 367, L9

\bibitem{} Heckman T.M., Armus L., Miley G.K., 1990, ApJS, 74, 833

\bibitem{} Heckman T.~M., Dahlem M., Eales S.~A., Fabbiano G., Weaver K., 1996, ApJ, 457, 616

\bibitem{} Hoopes C.~G., Heckman T.~M., Strickland D.~K., Howk J.~C., 2003, ApJ, 596, L175

\bibitem{} Houck J.C., Bregman J.N., Chevalier R.A., Tomisaka K., 1998, ApJ, 493, 431

\bibitem{} Ikebe Y., Leighly K., Tanaka Y., Nakagawa T., Terashima Y., Komossa S., 2000, MNRAS, 316, 433

\bibitem{} Iwasawa K., Fabian A.~C., Matt G., 1997, MNRAS, 289, 443

\bibitem{} Iwasawa K., Comsatri A., 1998, MNRAS, 297, 1219

\bibitem{} Iwasawa K., 1999, MNRAS, 302, 96

\bibitem{} Iwasawa K., Matt G., Guainazzi M., Fabian A.~C., 2001, MNRAS, 326, 894 

\bibitem{} Iwasawa K., Wilson A.S., Fabian A.C., Young A.J., 2003, 345, 369

\bibitem{} Joseph, R.D., Wright, G.S., 1985, MNRAS, 214, 87

\bibitem{} Kaastra J.S., 1992, An X-Ray Spectral Code for Optically Thin Plasmas, Internal SRON-Leiden Report, updated version 2.0

\bibitem{} Kallman T., Bautista M., 2001, ApJS, 133, 221

\bibitem{} Kaspi S., et al., 2002, ApJ, 574, 643

\bibitem{} Kennicutt R.C.Jr, 1998, ARAA, 36, 189

\bibitem{} Kinkhabwala A., et al., 2002, ApJ, 575, 732

\bibitem{} Komossa S., Burwitz V., Hasinger G., Predehl P., Kaastra J.S., Ikebe Y., 2003, ApJ, 582, 15

\bibitem{} Krolik J.~H., Madau P., \.Zycki P.~T., 1994, ApJ, 420, L57 

\bibitem{} Krolik J.H., Kriss G.A., 2001, ApJ, 561, 684

\bibitem{} Leitherer C., et al, 1999, ApJS, 123, 3

\bibitem{} Maiolino R., et al, 2003, MNRAS, 344, 59

\bibitem{} Matt G., 2002, MNRAS, 337, 147

\bibitem{} McDowell J.C., et al., 2003, ApJ, 591, 154

%\bibitem{} Mihos J.C., Hernquist L, 1994, ApJ, 431, L9

\bibitem{} Moran, E.C., Lehnert M.D., 1997, ApJ, 478, 172

\bibitem{} Nandra K., Mushotzky R.F., Arnaud K., Steidel C.C., Adelberger K.L., Gardner J.P., Teplitz J.I., Windhorst R.A., 2002, ApJ, 576, 625

\bibitem{} Norris, R.P., 1988, MNRAS, 230, 345

\bibitem{} Persic M., Rephaeli Y., 2002, AA 382, 843

\bibitem{} Persic M., Rephaeli Y., Braito V., Cappi M., Della Ceca R., Franceschini A., Gruber D.E., 2004, A\& A, 419, 849

\bibitem{} Pietsch W., et al., 2001, A\&A, 365, L174

\bibitem{} Ranalli P., Comastri A., Setti G., 2003, A\&A, 399, 39

\bibitem{} Richstone D., et al 1998, Nat, 395, 14

\bibitem{} Rieke G.H., 1988, ApJ, 331, L5

\bibitem{} Sakamoto K., Scoville N.Z., Yun M.S., Crosas M., Genzel R., Tacconi L.J., 1999, ApJ, 514, 68

\bibitem{} Sanders, D.B., Mazzarella, J.M., Kim, D.-C., Surace, J.A.,  2003, AJ, 126, 1607

\bibitem{} Sanders D.B., Scoville N.Z., Soifert B.T., 1991, ApJ, 370, 158

\bibitem{} Sanders D.B., Soifer B.T., Elias J.H., Madore B.F., Matthews K., Neugebauer G., Scoville N.Z., 1988, ApJ, 325, 74

\bibitem{} Scoville N.Z., Evans A.S., Thompson R., Rieke M., Hines D.C., Low F.J., Dinshaw N., Surace J.A., Armus L., 2000, AJ, 119, 991

\bibitem{} Scoville, N.Z., Sanders, B.T., Sargent, A.I., Soifer, B.T., Scott, S.L., Lo, K.Y., 1986, ApJ, 311, L47

\bibitem{} Scoville N.Z., Yun M.S., Bryant P.M., 1997, ApJ, 484, 702

\bibitem{} Scoville N.Z., et al., 1998, ApJ, 492, L107

\bibitem{} Smith H.E., Lonsdale C.J., Lonsdale C.J., Diamond P.J., 1998, ApJ, 493, L17
  
\bibitem{} Smith H.E., Lonsdale C.J., Lonsdale C.J., Diamond P.J., 1999, Ap\& SS, in ``Ultraluminous Infrared Galaxies: Babies or Monsters'', eds. R. Genzel, D. Lutz \& L. Tacconi, 266, 125

\bibitem{} Soifer, B.T., Neugebauer, G., Helou, G., Lonsdale, C.J.,  Hacking, P., Rice, W., Houck, J.R., Low, F.J., Rowan-Robinson, M., 1984, ApJ, 283, L1

\bibitem{} Soifer B.T., Neugebauer G., Matthews K., Egami E., Weinberger A.J., 2002, AJ, 124, 2980

\bibitem{} Solomon P.M., Radford S.J.E., Downes D., 1990, ApJ, 348, L53

\bibitem{} Spoon H.W.W., Keane J.V., Tielens A.G.G.M., Lutz D., Moorwood A.F.M., 2001, A\&A, 365, L353

\bibitem{} Spoon H.W.W., Keane J.V., Tielens A.G.G.M., Lutz D., Moorwood A.F.M., Laurent O., 2002, A\&A, 385, 1022

\bibitem{} Spoon H.W.W., Moorwood A.F.M., Lutz D., Tielens A.G.G.M., Seibenmorgen R., Keane J.V., 2004, A\&A, 414, 873


\bibitem{} Suchkov A.~A., Balsara D.~S., Heckman T.~M., Leitherner C., 1994, ApJ, 430, 511 

\bibitem{} Suchkov A.A., Berman V.G., Heckman T.M., Balsara D.S., 1996, ApJ, 463, 528

\bibitem{} Strickland D.K., Stevens I.R., 2000, MNRAS, 314, 511

\bibitem{} Strickland D., 2002, Chemical Enrichment of the Intracluster and Intergalactic Medium'', ASP Conference Proceedings Vol 253, eds. R. Fusco-Femiano and F. Matteucci, p387
 
\bibitem{} Strum E., et al, 1996, A\&A, 315, L133

\bibitem{} Taniguchi Y., Yoshino A., Ohyama Y., Nishiura S., 1999, ApJ, 514, 660

\bibitem{} Tomisaka K., Ikeuchi S., 1988, 330, 695

\bibitem{} Veilleux S., Kim D.-C., Sanders D.B., 1999, ApJ, 522, 113

\bibitem{} Weaver K.A., Heckman T.M., Strickland D.K., Dahlem M., 2002, ApJ, 576, L19

\bibitem{} Wright, G.S., James, P.A., Joseph, R.D., McLean, I.S., 1990, Nature, 344, 417

\bibitem{} Young A.J., Wilson A.S., Shopbell P.L., 2001, ApJ, 556, 6

\end{thebibliography}
\end{document}